\documentstyle[12pt]{article}
\begin{document}

\title{Time Dependent Partial Waves and Vortex Rings in the
 Dynamics of Wave Packets\footnote{submitted to J.~Phys.~A}}
\author{R.~Arvieu$^a$, P.~Rozmej$^b$ and
W.~Berej$^b$ \\[3mm] 
$^a$ Institut des Sciences Nucl\'eaires, F 38026 Grenoble-Cedex, France\\
arvieu@frcpn11.in2p3.fr\\
$^b$ Theoretical Physics Department, University MCS, 20-031 Lublin, Poland\\
rozmej@tytan.umcs.lublin.pl}
\date{\normalsize February 13, 1997}

\maketitle

\begin{abstract}
We have found a new class of time dependent partial waves which are
solutions of time dependent Schr\"odinger equation for three dimensional
harmonic oscillator. We also showed the decomposition of coherent states
of harmonic oscillator into these partial waves.
This decomposition appears particularly convenient for a description
of the dynamics of a wave packet representing a particle with spin
when the spin--orbit interaction is present in the hamiltonian.
An example of an evolution of a localized wave packet into a torus and
backwards, for a particular initial conditions is analysed
in analytical terms and shown with a computer graphics.

\vspace*{2mm}\noindent
PACS number(s): 03.65.Ge, 32.90+a
\end{abstract}

\normalsize

\section{Introduction}

The rapid technological development of short--pulsed lasers during
the last decade made it possible to produce and detect particular
states,
coherent superpositions of stationary electron states for a wide variety
of physical systems. The dynamics of these initially well localized
wave packets is a subject of much current investigation in many areas
of physics and chemistry \cite{alber,garraway}.
An extensive research by means of both theoretical and experimental
methods has brought an understanding of the intriguing phenomena
 of a hierarchy of collapses and revivals for particularly prepared
wave packets in Jaynes--Cummings Model (JCM) of quantum optics
$[$3--7$]$     and
in Rydberg atoms 
$[$8--14$]$.

In present paper we construct new solutions of time dependent
Schr\"odinger equation for spherical harmonic oscillator (HO). We
use these solutions further for description of the evolution
of wave packets representing a particle with spin moving in HO potential
with additional spin--orbit interaction. Time dependent partial waves
allow for much clearer interpretation of subtle interference
effects as well as for a substantial acceleration of numerical codes
used for graphical presentation of a complicated wave packet motion.

The paper is organized as follows.
In section 2 we construct time dependent partial waves for the
harmonic oscillator and show the decomposition of coherent states
into these states. In section 3 we show the motion of individual
partial waves. In section 4 we discuss time dependent spinors
corresponding to solutions of time dependent Schr\"odinger equation
for hamiltonian containing spin--orbit interaction and show a
possibility of evolution of an initially localized wave packet into
a toroidal shape and backwards. Sections 5 and 6 present graphical 
illustrations of the wave packet motion and conclusions.

\section{Time dependent partial waves for the harmonic oscillator}

It is not so well known that there exist simple time dependent
partial waves which are solutions of the Schr\"odinger equation
of the three dimensional oscillator. The proof will be given
below.

We look for solutions $\psi^m_l(\vec{r},t)$ of the time dependent
Schr\"odinger equation for the spherical harmonic oscillator
(with $\hbar=m=\omega=1$) with the following form
\begin{equation}
  \psi^m_l(\vec{r},t) = F(t)\,e^{-\frac{1}{2}r^2}\, W_l(R_0e^{-i t}r)
 \, Y^m_l(\theta,\varphi )
\label{equ1}
\end{equation}
where $F(t)$ and $W_l$ must be determined in terms of a complex
number $R_0$.

We find easily that $W_l$ must obey the differential equation
\begin{equation}
 Z^2\,\frac{d^2 W_l}{d Z^2} +2\,Z\,\frac{d W_l}{d Z}
 -[l(l+1)+(3-\frac{2i }{F}\,\frac{d F}{d t})\,r^2]\, W_l = 0
\label{equ2}
\end{equation}
with a new variable $Z$
\begin{equation}
 Z = R_0 \, e^{-i t} r\,.
\label{equ3}
\end{equation}
$W_l$ depends only on this variable if $F$ solves the equation
\begin{equation}
 3-\frac{2i }{F}\,\frac{d F}{d t} = {R_0}^2\, e^{-2i t} \;.
\label{equ4}
\end{equation}
Within an arbitrary constant factor, which is not written the
solution is
\begin{equation}
 F(t) = e^{-\frac{3}{2}i t}\,e^{-\frac{1}{4}{R_0}^2\,e^{-2i t}} \;.
\label{equ5}
\end{equation}
The equation solved by $W_l$ becomes
\begin{equation}
 Z^2\,\frac{d^2 W_l}{d Z^2} +2\,Z\,\frac{d W_l}{d Z}
 -[l(l+1) + Z^2]\,W_l = 0
\label{equ6}
\end{equation}
which is the equation for the modified spherical Bessel
functions.

In the following we will use only the modified spherical Bessel
functions of the first kind with the usual conventions of the
literature \cite{abram}
\begin{eqnarray}
 W_l(Z) & = & \sqrt{\frac{\pi}{2Z}}\,I_{l+\frac{1}{2}}(Z) \\
	& = & \frac{Z^l}{(2l+1)!!} \left[
   1 + \frac{\frac{1}{2}Z^2}{1!\,(2l+3)} +
       \frac{(\frac{1}{2}Z^2)^2}{2!\,(2l+3)(2l+5)} + \cdots \right] \;.
\label{equ7}
\end{eqnarray}
The interpretation of these waves is very simple: there is
an harmonic motion in the radial part of the wave function
keeping the angular part the same because of angular momentum
conservation.

The partial waves $\psi^m_l(\vec{r},t)$ 
just defined occur in a~natural way in the
expansion of a~coherent time dependent gaussian wave packet
into partial waves. In such a~wave packet the constant $R_0$
finds its interpretation by combining the position of the
center of the wave packet $\vec r_0$ to its mean momentum
$\vec p_0$ at a~time $t=0$. Let $\vec{r}_t$ and $\vec{p}_t$ be such
vectors at time $t$ and let us define
\begin{equation}
  \vec{R}_t = \vec{r}_t + i \vec{p}_t \;,
\label{equ9}
\end{equation}
while
\begin{equation}
  \vec{R}_0 = \vec{r}_0 + i \vec{p}_0 \;.
\label{equ10}
\end{equation}
These two vectors are related by
\begin{equation}
  \vec{R}_t = \vec{R}_0\,e^{-i t} \;.
\label{equ11}
\end{equation}
A normalized gaussian wave packet centered on $\vec{r}_t$ with
mean momentum $\vec{p}_t$ is now written as
\begin{equation}
 \bar{\psi}_{\vec{R}_t}(\vec{r}) = \pi^{-\frac{3}{4}}\,
 e^{-\frac{1}{2}(\vec{r}-\vec{r}_t)^2}\,e^{i \vec{p}_t\cdot\vec{r}}
 = \pi^{-\frac{3}{4}}\,
e^{-\frac{1}{2}r^2}\,e^{-\frac{1}{2}{r_t}^2}\,e^{\vec{r}\cdot\vec{R}_t}
\;.
\label{equ12}
\end{equation}
Using 10.2.36 from reference \cite{abram} the modified Bessel function
of the
first kind of argument $Z = R_t\, r=R_0\, e^{-i t}\,r$ appears in the
expansion of the last exponential of (12)
\begin{equation}
 e^{\vec{r}\cdot\vec{R}_t} = \sum_{l=0}^{\infty}\, (2l+1)\,
 \sqrt{\frac{\pi}{2Z}}\,I_{l+\frac{1}{2}}(Z)\, P_l(\cos \Theta)
\label{equ13}
\end{equation}
and $R_0$ is defined as
\begin{equation}
  R_0 = (\vec{R}_0 \cdot \vec{R}_0)^{\frac{1}{2}}
  = ({r_0}^2 - {p_0}^2 + 2i \vec{r}_0\cdot\vec{p}_0)^{\frac{1}{2}} \;.
\label{equ14}
\end{equation}
In a gaussian wave packet all the time dependent partial waves
share the same parameter $R_0$. However it is possible to
consider more general wave packet for which $R_0$ might be
different for different $l$.

The Legendre polynomial $P_l(\cos\Theta)$ depends on the
(complex) angle $\Theta$ between $\vec r$ and $\vec R_t$.
However this angle is time independent. Indeed for each
direction $\alpha$, $\alpha=x,y,z$ one has
\begin{equation}
 r_{\alpha}(t) +i p_{\alpha}(t) = (r_{\alpha}(0) +i p_{\alpha}(0))
\, e^{-i t}
\label{equ15}
\end{equation}
as well as equation (11), therefore
\begin{eqnarray}
 \cos(\Theta) & = &
 \frac{x[x(t)+i p_x(t)]+y[y(t)+i p_y(t)]+z[z(t)+i p_z(t)]}
 {R_t\, r} \\
 & = & \frac{x[x(0)+i p_x(0)]+y[y(0)+i p_y(0)]+z[z(0)+i p_z(0)]}
 {R_0\, r} \;.
 \nonumber
\label{equ16}
\end{eqnarray}
Then if the complex direction $\theta_{R_0}$, $\varphi_{R_0}$ of
$\vec R_0$ is introduced through
\begin{eqnarray}
  \cos\theta_{R_0}~~~~~~~~~~ & = & \frac{z(0)+i p_z(0)}{R_0} \nonumber
  \\
  \sin\theta_{R_0}\,\cos\varphi_{R_0} & = & \frac{x(0)+i p_x(0)}{R_0}
  \\
  \sin\theta_{R_0}\,\sin\varphi_{R_0} & = & \frac{y(0)+i p_y(0)}{R_0}
  \nonumber
\label{equ17}
\end{eqnarray}
one writes the addition formula:
\begin{equation}
 P_l(\cos\Theta) = \frac{4\pi}{2l+1} \,\sum_{m=-l}^{l} \, (-1)^m\,
 Y^m_l(\theta,\varphi)\,
 Y^{-m}_l (\theta_{R_0}, \varphi_{R_0}) \;.
\label{equ18}
\end{equation}
We can now rewrite $\bar{\psi}_{\vec R_t}$ as
\begin{equation}
  \bar{\psi}_{\vec{R}_t}(\vec{r}) = 4 \pi^{\frac{1}{4}}\,
 e^{-\frac{1}{2}(r^2+r_t^2)}\,\sum_{l=0}^{\infty} \, \sum_{m=-l}^{l}\, 
 (-1)^m\,  Y^{-m}_l (\theta_{R_0}, \varphi_{R_0})
 W_l(R_t r)\,Y^m_l(\theta,\varphi) \;.
\label{equ19}
\end{equation}
We will now  extract the function $F$ (5) in order to
show explicitly the time dependent partial waves
\begin{eqnarray}
 e^{-\frac{1}{4}R_0^2\,e^{-2i t}} & = &	e^{-\frac{1}{4}
 (r_t^2-p_t^2+2i \vec{r}_t\cdot\vec{p}_t)} \\  & = &
 e^{-\frac{1}{2}r_t^2} \, e^{-\frac{i }{2}\vec{r}_t\cdot\vec{p}_t}\,
 e^{\frac{E_0}{2}} \;.	\nonumber
\label{equ20}
\end{eqnarray}
The classical energy $E_0=\frac{r^2_0+p^2_0}{2}$ has been
introduced above. The time dependent partial waves (1) which
appear in the gaussian wave packet at time $t$ are then
expressed as:
\begin{equation}
 \psi^m_l(\vec{r},t) =
e^{-(\frac{3}{2}i t + \frac{i }{2}\vec{r}_t\cdot\vec{p}_t
 + \frac{r^2+r_t^2}{2} -\frac{E_0}{2})}\, W_l(R_t r)\,
 Y^m_l(\theta,\varphi) \;.
\label{equ21}
\end{equation}
The gaussian wave packet (19) can now be expressed as follows in
terms of (21)
\begin{equation}
 \bar{\psi}_{\vec R_t}(\vec{r}) = 4\pi^{\frac{1}{4}}\,
 e^{i (\frac{3}{2}t +
 \frac{\vec{r}_t\cdot\vec{p}_t}{2})-\frac{E_0}{2}}
 \sum_{lm}\, (-1)^m\,  Y^{-m}_l (\theta_{R_0}, \varphi_{R_0})
 \psi^m_l(\vec{r},t) \;.
\label{equ22}
\end{equation}
Clearly one obtains a solution of the Schr\"odinger equation
only if we change the phase. Let $\psi_{\vec{R}_t}(\vec{r})$ be
this solution
\begin{eqnarray}
 \psi_{\vec{R}_t}(\vec{r}) & = &
 e^{-i (\frac{3}{2}t + \frac{\vec{r}_t\cdot\vec{p}_t}{2})}\,
 \bar{\psi}_{\vec R_t}(\vec{r}) \nonumber \\
 & = &	 \sum_{lm}\, (-1)^m\, 4\pi^{\frac{1}{4}}\, e^{-\frac{E_0}{2}}
Y^{-m}_l (\theta_{R_0}, \varphi_{R_0}) \,
 \psi^m_l(\vec{r},t)	   \\
& = &	\sum_{lm}\,C_{lm}(\vec R_0)  \psi^m_l(\vec{r},t) \;.\nonumber
\label{equ23}
\end{eqnarray}
The weight $C_{l_m}(\vec R_0)$ of the time dependent partial
wave $\psi^m_l(\vec{r},t)$ is now seen to be
\begin{equation}
 C_{lm}(\vec{R}_0) = (-1)^m \, 4\pi^{\frac{1}{4}}\, e^{-\frac{E_0}{2}}
 \, Y^{-m}_l (\theta_{R_0}, \varphi_{R_0}) \;.
\label{equ24}
\end{equation}

An expansion similar to equation (23) is found in reference
\cite{boris}.
However the time dependence of the radial function is
lacking in this reference as well as the phase which
is needed to correct the gaussian wave packet. The existence
of this phase is mentioned in some textbooks \cite{textbk}.

\section{Representation of the time dependent partial waves}

Let us define the phase of $R_0$ by
\begin{equation}
 R_0 = ({r_0}^2-{p_0}^2+2i \vec{r}_0\cdot\vec{p}_0)^{\frac{1}{2}}
 = |R_0|\,e^{i \delta_0} \;.
\label{equ25}
\end{equation}
The argument of the modified Bessel function is 
$Z=r|R_0|e^{-i (t-\delta_0)}$.
Changing the initial conditions of the wave packet produces 
a~change of $|R_0|$ and of $\delta_0$. However it is clear
that the change of the phase can be taken into account in the
radial partial wave $W_l(z)$ by a~shift of the origin of time.
It is therefore possible to analyse the different partial waves
by assuming that $\delta_0=0$. This possibility arises
if $p_0=0$, $R_0=r_0$. For these values the gaussian wave packet
performs a~linear motion. It is striking that the radial motions
of the partial waves corresponding to such a~case contain all the
proper information which can be used for all the different
trajectories, i.e. circular or elliptical. The differences
between the trajectories will be introduced by the coefficients
$C_{lm}(\vec{R}_0)$.

In the following figures we have chosen to study the radial
waves for the mean energy $E_0=N=20$ i.e.
$r_0=\sqrt{2E_0}=\sqrt{40}=R_0$.
The 8 lowest partial waves present in the development of
a~gaussian wave packet are represented in figure 1  $[$as
a~matter of fact we have represented the density multiplied by
$r^2]$ for a~time range going from $t=0$ to ${T\over 2}$ ($T$ is
the harmonic oscillator period). The densities exhibit the same
motion toward the origin and of course this radial motion is
symmetric with respect to $t={T\over 4}$. However there is
clearly an effect of the centrifugal barrier since the waves with
the highest $l$ are repelled from the origin. Also some secondary
maxima are present. It is worth to
mention once again that the effects of the centrifugal barrier cancel
when one adds the waves to produce a~linear motion.
We should also note that the spread of the partial waves does
not depend on $l$ in a~significant manner.

In figure 2 the squares of 16 lowest partial waves ($|r\psi_l(r,t)|^2$) 
are represented
as functions of $r$ for $t=0$. The concentration of the
waves at the same $r$ is spectacular. The intensity of the
waves increases when $l$ goes from $l=0$ to $l=4$ and decrease
afterwards. In order to explain the concentration of the wave
in three dimensions the angular part plays a~role as will be
discussed later on.

\section{Time dependent spinors}

We will now study the evolution of a coherent wave packet which
is initially in a~pure spin eigenstate chosen as $s_z = +{1\over
2}$. This wave packet will evolve on the action of the harmonic
oscillator evolution operator $U_0(t)$ and of the spin--orbit
part $U_{ls}(t)$. As in our previous papers 
$[$18--22$]$
we will use the
spin--orbit hamiltonian with a~constant factor $\kappa$:
\begin{equation}
 V_{ls} = \kappa \, (\vec{l}\cdot\vec{\sigma})
\label{equ26}
\end{equation}
and therefore $U_0$ commute with $V_{ls}$. The hamiltonian
contains the ratio of the two time scales (let us recall that
$\omega=1$)
\begin{equation}
 \frac{T_{ls}}{T} = \frac{2\pi}{\kappa}\, \frac{\omega}{2\pi} =
 \frac{1}{\kappa}  \; .
\label{equ27}
\end{equation}
The fully normalized spinor is written at time $t=0$ as
\begin{equation}
 \stackrel{\sim}{\psi}_{\vec{R}_0}(\vec{r}) = \left(
 \begin{array}{c} \psi_{\vec{R}_0}(\vec{r}) \\ 0 \end{array} \right)
= \sum_{lm} \, C_{lm}(\vec{R}_0)\, \left(
 \begin{array}{c} \psi_l^m(\vec{r},0) \\ 0 \end{array} \right) \;.
\label{equ28}
\end{equation}
We have shown previously [4] that
\begin{equation}
 U_{ls}(t) = f(t) + g(t)\,(\vec{l}\cdot\vec{\sigma})
\label{equ29}
\end{equation}
with the following operators
\begin{eqnarray}
\label{equ30}
 f(t) & = & e^{i \frac{t}{2}}\left( \cos\Omega\frac{t}{2}-\frac{\i
 }{\Omega}
 \sin\Omega\frac{t}{2} \right) \\
\label{equ31}
 g(t) & = & e^{i \frac{t}{2}}\left( -\frac{2i }{\Omega}
 \sin\Omega\frac{t}{2} \right) \\
\label{equ32}
 \Omega|lm\rangle & = & \sqrt{1+4\hat{l}^2}\: |lm\rangle = (2l+1)\,
 |lm\rangle \;.
\end{eqnarray}
$U_0(t)$ transforms $\vec R_0$ into $\vec R_t$ while $U_{ls}(t)$
acts on the spherical harmonics present in the $\psi^m_l(\vec
r,t)$. The result is

\begin{eqnarray}
 \stackrel{\sim}{\psi}_{\vec{R}_t}(\vec{r}) & = &
e^{[-i (\frac{3}{2}t + \frac{\vec{r}_t\cdot\vec{p}_t}{2})
 - \frac{r^2+r_t^2}{2} -\frac{E_0}{2}]} \\
 & \times & \sum_{lm}\, C_{lm}(\vec{R}_0)\,
W_l(R_t r)\, \left( \begin{array}{r}
(f_l+mg_l)\,Y^m_l(\theta,\varphi) \\
\frac{1}{2}\,g_l\, \sqrt{l(l+1)-m(m+1)}\: Y^{m+1}_l(\theta,\varphi)
\end{array} \right) \;,  \nonumber
\label{equ33}
\end{eqnarray}
where $f_l(t)$ and $g_l(t)$ are simply obtained from (\ref{equ30})
and (\ref{equ31}) 
by replacing $\Omega$ by $(2l+1)$.
This expression corresponds to the most general initial
condition with the assumed initial spin direction. The presence
of the time dependence in the functions $f$ and $g$ from one
hand, the presence of $Y^{m+1}_l$ in the lower part of the
spinor are the two differences which change in an appreciable
manner the time evolution in the case where $\kappa \neq 0$.

We want now to concentrate our efforts on the evolution of a~wave
packet which is cylindrically symmetric around $Oz$ for $t=0$.
We assume therefore (denoting unit vector in $Oz$ direction by
$\hat{z}$)
\begin{eqnarray}
 \vec{r}_0 = -r_0\,\hat{z} \hspace{2ex},\hspace{2ex} \vec{p}_0 = 0
 \hspace{2ex},\hspace{2ex} R_0 = r_0 \hspace{2ex},\hspace{2ex} \\
 E_0=\frac{r_0^2}{2} \hspace{2ex},\hspace{2ex} \theta_{R_0}=\pi
 \hspace{2ex},\hspace{2ex} \varphi_{R_0}=0  \hspace{2ex},\hspace{2ex}
\label{equ34}
\end{eqnarray}
\begin{equation}
 C_{lm}(\vec{R}_0) =  \delta_{m0}\,(-1)^l\, 2\pi^{-\frac{1}{4}}\,
 e^{-\frac{N}{2}}\, \sqrt{2l+1} \;.
\label{equ36}
\end{equation}
Let us study the sign of the product $C_{l0}(\vec
R_0)\,Y^0_l(\theta,\varphi)$ along the $Oz$ axis i.e.
$\theta=\pi$ and  $\theta=0$. One has
\begin{eqnarray}
  Y^0_l(\pi,0) & = & (-1)^l \, \sqrt{\frac{2l+1}{4\pi}}  \\
  Y^0_l(0,0) & = & ~~~~~~~~\sqrt{\frac{2l+1}{4\pi}}  \;.
\label{equ37}
\end{eqnarray}
All the partial waves act coherently in the direction
$\theta=\pi$ while there is a~destructive interference for
$\theta=0$. We also know that there is a~radial concentration of
the wave packets shown on figure 2. This explains mainly the known
result that the wave packet is initially a~gaussian centered at
$\vec{r}_0=-r_0\,\hat{z}$.

In the direction $\theta={\pi\over 2}$ we have
\begin{equation}
  Y^0_l(\frac{\pi}{2},0) = 0 \hspace{3ex} {\mathrm ~if~} l {\mathrm
  ~is~odd}
\label{equ39}
\end{equation}
while for $l$ even its sign is $(-1)^{l\over 2}$. A~destructive
interference between the even states is also predicted on the
$xOy$ plane with a~different character as for the $+Oz$ direction.

We will now try to repeat these arguments for $t\neq 0$ for each
component of the spinor (33).
For the part with $s_z +{1\over 2}$ we must find the phase of
the product $C_{l0}\,Y^0_l(\theta,\varphi)\,f_l(t)$. At a~time
$t={T_{ls}\over 4}={\pi\over 2}$ (assuming $\kappa=1$) we have
for high $l$:
\begin{equation}
 f_l(\frac{\pi}{2}) \approx e^{i \frac{\pi}{4}}\,
 \cos(2l+1)\frac{\pi}{4}
 = e^{i \frac{\pi}{4}}\, \frac{\sqrt{2}}{2}\,(-1)^\frac{l}{2} \; .
\label{equ40}
\end{equation}
The coherence of the product considered is now obtained for
$\theta={\pi\over 2}$ and for even $l$. It is interesting to
see that the part with $s_z=+{1\over 2}$ has still cylindrical
symmetry around $Oz$ but moreover if $t={T_{ls}\over 4}$ or $0$ we
shall have the result of figure 2 i.e. a~high radial concentration
of the wave. One readily understand that at this time the wave
is highly peaked on a~vortex of radius $r_0$ with symmetry
around $Oz$, the smaller radius of the vortex being roughly the
initial radial spread.

As for the part with $s_z=-{1\over 2}$ we must discuss the sign
of the product $C_{l0}\,Y^1_l(\theta,\varphi)\,g_l(t)$ $\times \,
\sqrt{l(l+1)}$ also for $t={\pi\over 2}$ and for
$\theta={\pi\over 2}$. Now $Y^1_l({\pi\over 2},\varphi)=0$
if $l$ is even, its sign for odd $l$ is $(-1)^{l-1\over 2}$ while
$g_l({\pi\over 2})$ has also this sign since
\begin{equation}
 g_l(\frac{\pi}{2}) = -e^{i \frac{\pi}{4}}\, \frac{2i }{2l+1}\,
 \sin(2l+1)\frac{\pi}{4} = \frac{-2i }{2l+1}\, (-1)^{l-1\over 2} \,
 e^{i \frac{\pi}{4}}\, \frac{\sqrt{2}}{2} \; .
\label{equ41}
\end{equation}
However we have also $Y^1_l(\theta,\varphi)\approx
e^{i\varphi}$. In the case of the part with $s_z=-{1\over
2}$ it is the odd partial waves which produce a~vortex similar
to that to the part with $s_z=+{1\over 2}$ in the limit of high
$l$. A~difference occurs in a $\varphi$ dependent phase.

For intermediate values of the time it is not possible to
provide a~similar discussion. However, by continuity we can
understand that the vortex rings are created from the spherical
initial wave packet and get their full extension for the
configuration and at the time chosen.

\section{Numerical simulation of the vortex rings}

The manifestation of the vortex rings depends on the parameter
$\kappa$. It is simpler to freeze the evolution under $U_0$ and to
consider only the spin--orbit evolution to begin with
in order to emphasize the effect. Figures 3 and 4 represent the
density of the full wave packet at various times on the planes
$yOz$ and $xOz$, respectively. Here also we have considered the wave
packet which was analyzed in figure 2.
$[$For convenience the spin direction is
$s_x=+{1\over 2}$ and $\vec{r}_0=r_0\,\hat{x}$ at time
$t=0]$. At time $t=0$ a~gaussian is represented.
At time $t={T_{ls}\over 8}$ a vortex ring has been created and gets
the maximum radius at $t={T_{ls}\over 4}$
when intersecting the plane $yOz$. Moreover
figure 4 shows that even a~second vortex with a~smaller amplitudes
has also been created that we cannot explain into simple terms.
At time $t={3T_{ls}\over 8}$ the vortex ring has a~decreasing
radius and is centered on a~point with $x<0$. Finally at time
$t={T_{ls}\over 2}$ the wave packet is reassembling near a~point
with $\vec{r}_0=-r_0\,\hat{x}$. At this time we have
shown in our previous papers that the spin is approximately
reversed. Finally the behaviour for $T_{ls}/2 < t < T_{ls}$ has
not been shown for it is totally reversible if we assume
a~frozen oscillator.

The vortex rings evolve approximately on a sphere. The
intersections of the wave packet with a~plane $xOz$ shown on
figure 5 present indeed the feature that the distance to the
center is about $r_0$ at all times.

The following figure 6 shows the decomposition of the toroidal wave
packet into its spin components in a direction perpendicular
to the classical motion. The left and central columns display 
the shapes of subpackets in which spin field is antiparallel and
parallel
to {em Ox} axis, respectively. Due to  symmetry with respect
to {em Oz} axis (classical trajectory), the similar decomposition 
with respect to spin components in an arbitrary direction perpendicular
to {em Oz} axis results in a picture rotated by a certain angle.

\section{Conclusions}

  The dynamics of wave packets in static potentials implies subtle
  interference 
effects which have been beautifully illustrated in the case of Coulomb 
potential $[$9-14,16$]$. Two general mechanisms have been identified: 
one is the spread of the wave on the top of a classical trajectory,
the other is the regime of partial and also almost complete recurrences.
In the case of a pure harmonic oscillator the existence of a single
frequency 
allows a unique interference mechanism of the partial wave  that leads
to 
a dispersionless coherent wave packet. 
This is the case where quantum mechanics comes closer to classical
mechanics:
the dynamics of the density probability is identical to the dynamics of 
the density distribution of the classical ensemble. 
This property is expected to be lost if a perturbation is added to the 
potential. If the perturbation is a spin--orbit potential the dynamics 
of the problem resembles very much that of the JCM \cite{shore}.  
There exist two versions \cite{phoen} 
of this model where the time evolution is exactly 
periodic in the closest analogy to our model: the Raman coupled model
and the 
two photons JCM. In the harmonic oscillator with spin--orbit the
dynamics is 
however richer if one follows the angular evolution with different
initial 
conditions. We have shown previously \cite{arvroz1} that the amount of
partial 
revival for half a spin--orbit period  depends indeed on these
conditions. 
We have discussed extensively \cite{arvroz,rozarv1} 
cases where the wave packet is divided 
into two parts which rotate in opposite directions on a circular average

trajectory in much the same way as the Stern--Gerlach effect. 
In the present paper we have shown a new dynamical behaviour for
cylindrically 
symmetric initial conditions: vortex rings are 
created and destroyed periodically. In a way this is a case where the 
coherence of the wave is kept for the $r$ and $\theta$ coordinates but
where 
the spread is applied only to the $\phi$ variable. 
It is an open question whether this effect relies only on the properties
of 
the partial waves of the harmonic oscillator discussed in this paper. 

{\sf Acknowledgements}

One of us (P R) likes to express his thanks to ISN, Grenoble for
kind hospitality during his stay in June--July 1996, when much of work
have been done.

\vspace*{8mm}
{\large\bf Figure captions}
 
\begin{description}
\item[Fig.~1.]
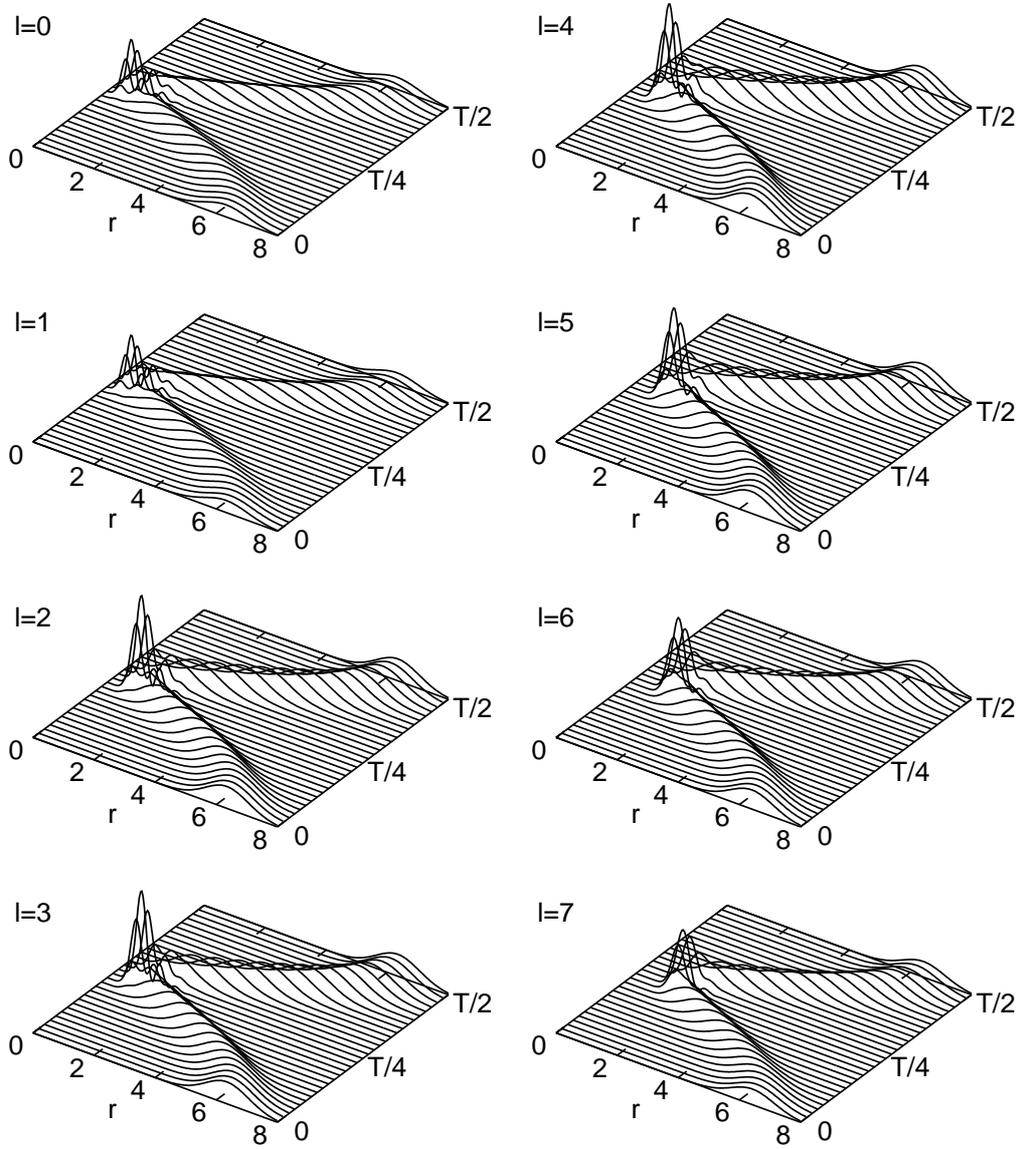
Radial motion of eight lowest partial partial waves with
$l=0,1,\dots,7$ contributing to the linear motion of the gaussian
wave packet. Shown is $|r\psi_l(r,t)|^2$ as a function of $r$ for
30 time steps in the interval $t\in [0,\frac{T}{2}]$.
The case $E_0=N=20$ is presented. The vertical scale is the same for
all figures.

\item[Fig.~2.]
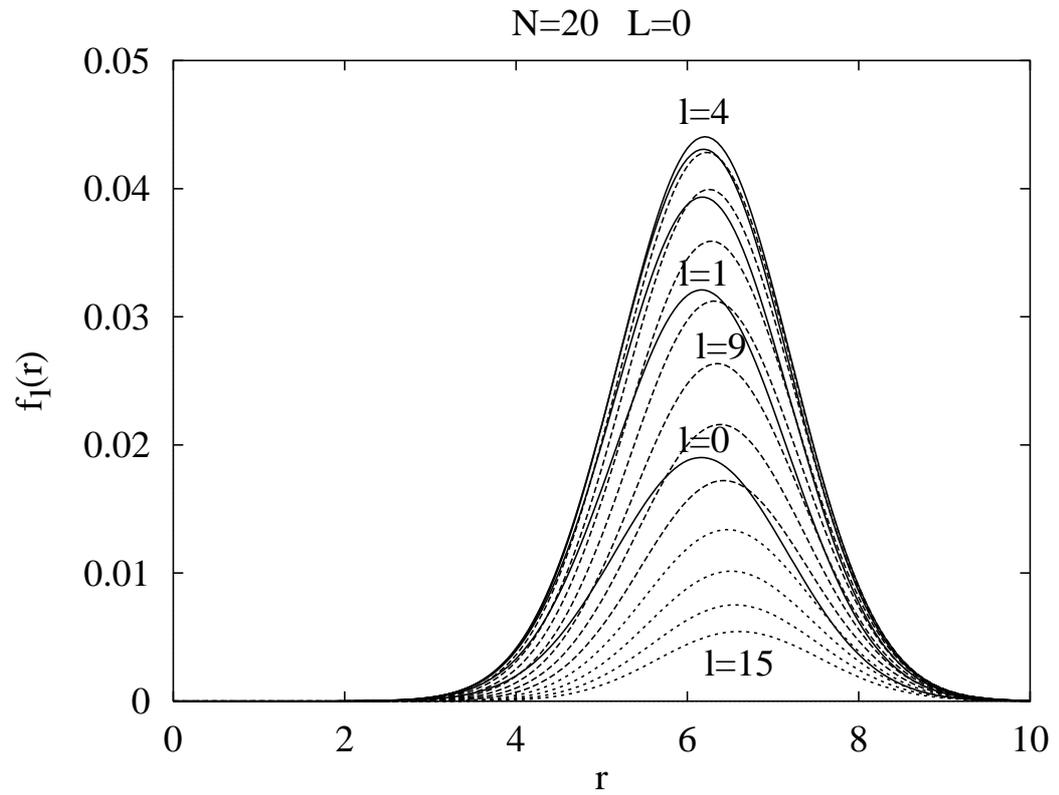
Decomposition of the gaussian wave packet into partial waves at
$t=0, r_0=\sqrt{2N}=\sqrt{40}$. The lowest 16 partial waves with
$l=0,1,\dots,15$ are shown. Waves with $l=0-4$ are represented by solid
lines (intensity increases with $l$),
those with $l=5 - 15$ by dashed ones (their intensities decrease).

\item[Fig.~3.]
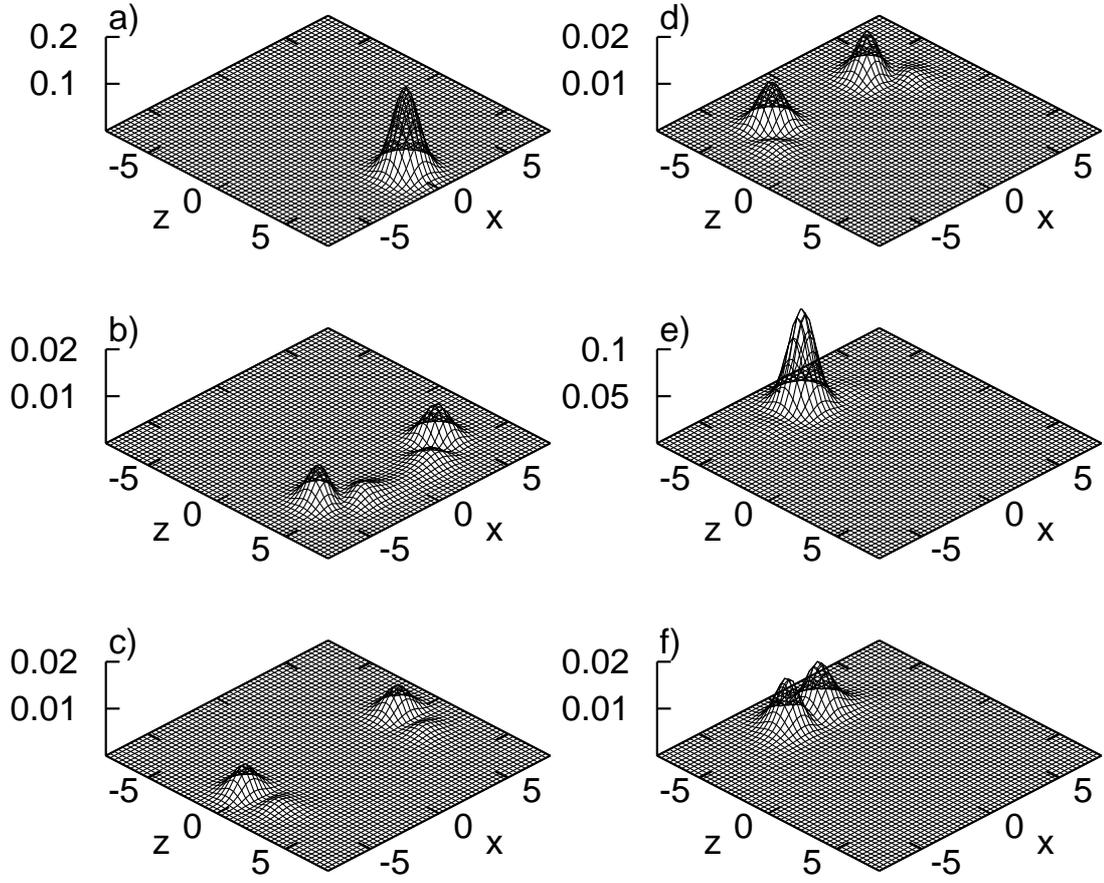
Time evolution of the wave packet with spin under $U_{ls}(t)$ operator
only (evolution according to $U_0$ is frozen).
Shown is the $|\Psi(t)|^2=|\Psi_+(t)|^2+|\Psi_-(t)|^2$ as a function
of coordinates on the plane $xOz$. Case $N=20$.
Note different vertical scales. Cases {em a,b,c,d,e,f} correspond to 
$t_i=0$, 1/8, 2/8, 3/8 15/32 and $4/8\,T_{ls}$ respectively.

\item[Fig.~4.]
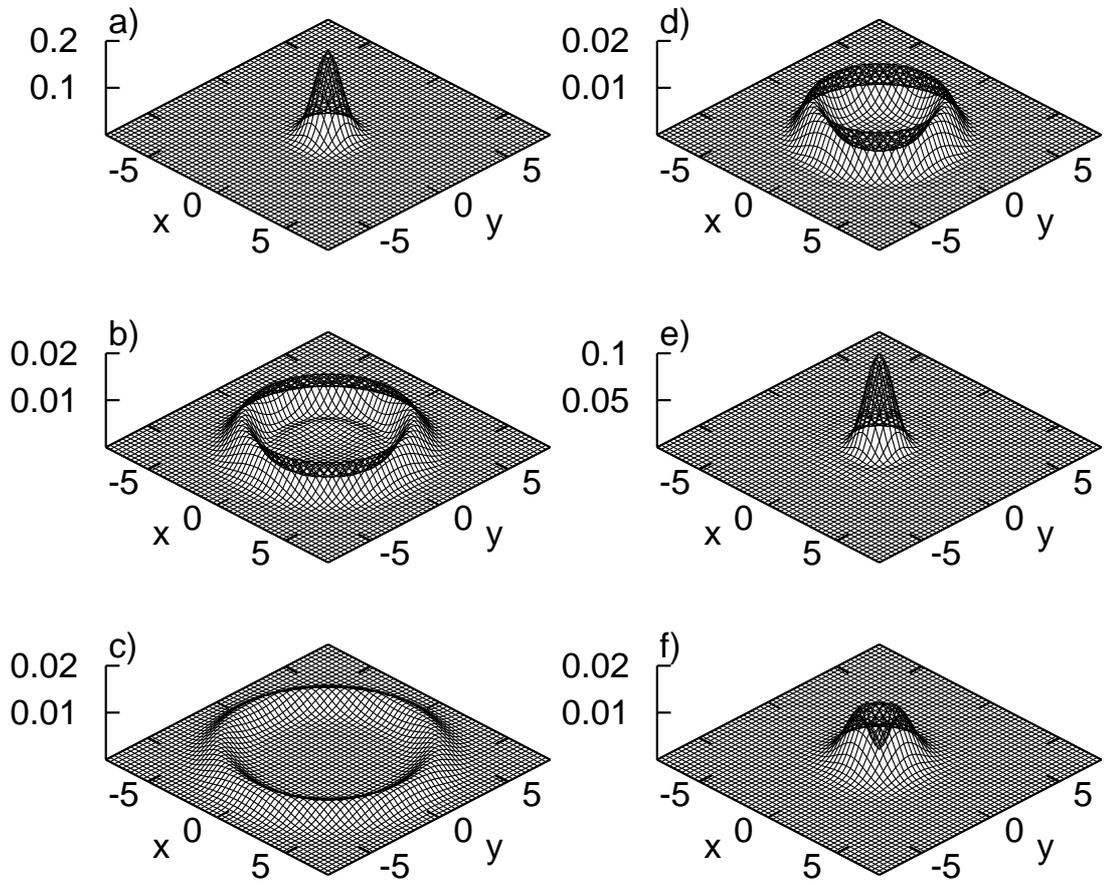
The same wave packet as in previous figure. Here shown are cuts
through planes perpendicular to the classical trajectory ({em Oz}
axis ) with $z_i=z_0\,\cos t_i$ showing explicitly a toroidal shape.
Time instants as in previous figure. 

\item[Fig.~5.]
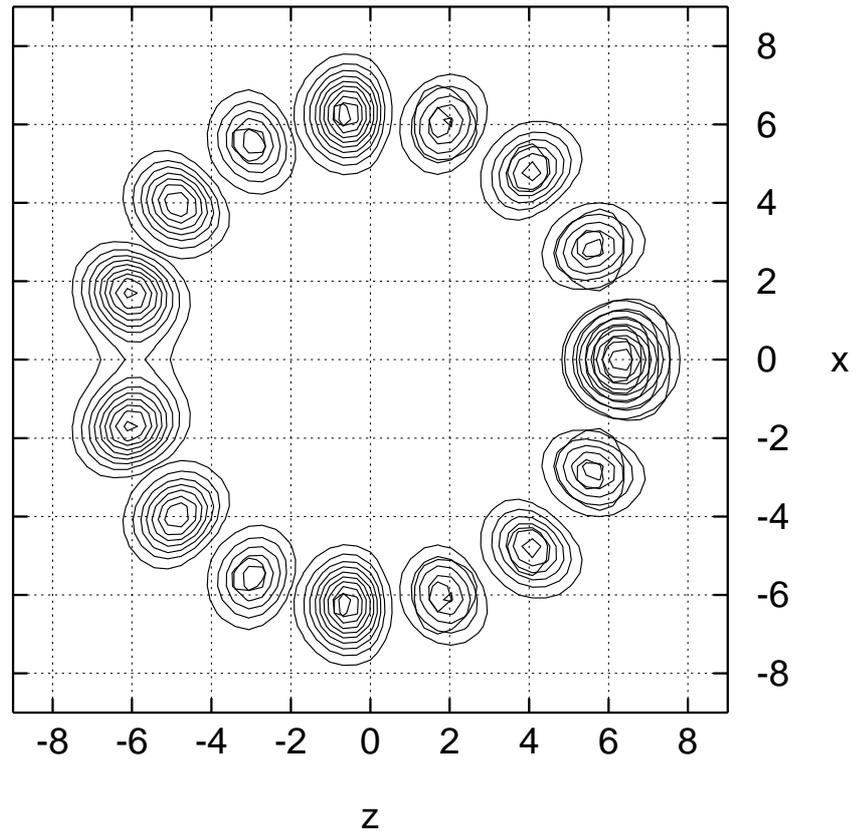
Contour plots of  the same evolution as in figure 4,  showing that
centers
of the vortex rings evolve approximately on a sphere with the radius
$r_0$.
Here packets corresponding to different time instants are collected in 
the same picture. 

\item[Fig.~6.]
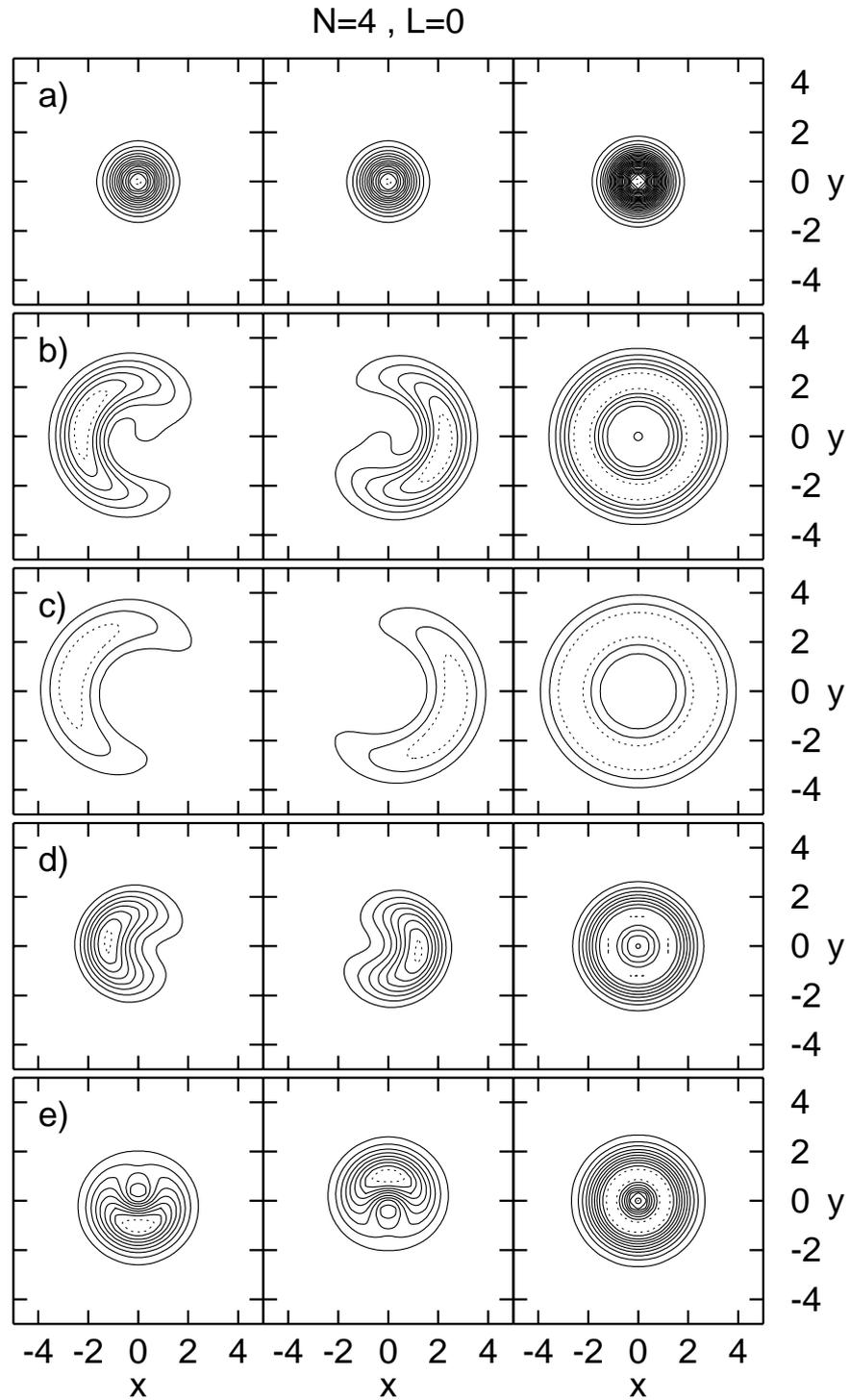
Contour plots showing the decomposition of the total wave packet
(right column) into parts with the opposite spin field (left and central
columns). Case $N=4$. Top row a) corresponds to $t=0$, bottom e)
to  $t=\frac{1}{2}T_{ls}$. Time steps are
$\Delta\,t=\frac{1}{8}T_{ls}$.
\end{description}


\begin{thebibliography}{99}
\bibitem{alber} Alber G and Zoller P 1991 {\it Phys.\ Rep.} {\bf 199}
231.
\bibitem{garraway} Garraway B M and Suominen K-A 1995  
 {\it Rep.\ Prog.\ Phys.} {\bf 58} 365.
\bibitem{jaynes} Jaynes E T and Cummings F W 1963 {em Proc. IEEE} {\bf
51} 129.
\bibitem{buck} Buck B and Sukumar C V 1981 {em Phys.~Lett.} A {\bf 81}
132.
\bibitem{kni} Knight P L 1986 {em Phys.~Scripta} T {\bf 12} 51.
\bibitem{gea} Gea--Banacloche J 1990 {em Phys.~Rev.~Lett.} {\bf 65};\\
   1991 {\it Phys.~Rev.} A {\bf 44} 5913;\\
   1992 {\it Opt.~Commun.} {\bf 88} 531.
\bibitem{aver2} Averbukh I S 1992 {\it Phys.~Rev.} A {\bf 46} R2205.
\bibitem{brown} Brown L S 1973 {\it Am.~J.~Phys.} {\bf 41} 525.
\bibitem{park} Parker J and Stroud C R Jr. 1986 {\it Phys.~Rev.~Lett.}
 {\bf 56} 716.
\bibitem{aver} Averbukh I S and Perelman N F 1989 
  {\it Phys.~Lett.} A {\bf 139} 449;\\
   1989 {\it Zh. Eksp. Teor. Fiz.} {\bf 96} 818 
  (1989 {\it Sov. Phys JETP \bf 69} 464);\\
  {1991 \it Usp. Fiz. Nauk} {\bf 161} 41 
  (1991 {\it Sov. Phys Usp. \bf 34} 572).
\bibitem{dacic} Da\u{c}ic-Gaeta Z and Stroud C R Jr. 1990
 {\it Phys.~Rev.} A {\bf 42} 6803.
\bibitem{peres} Peres A 1993 {\it Phys.~Rev.} A {\bf 47} 5196.
\bibitem{waals} Wals J Fielding H H and van Linden van den Heuvell H B
1995
  {\it Physica Scripta}  T {\bf 58} 62.
\bibitem{bluhm} Bluhm R and Kostelecky V A 1995 {\it Phys.~Lett.} A {\bf
200} 308.
\bibitem{abram} Abramowitz M and Stegun I A 1964 {\it Handbook of
  Mathematical Functions} (National Bureau of Standards) sect.10.2.
\bibitem{boris} Boris S D Brandt S Dahmen H D and Stroh T 1993
  {\it Phys.~Rev.}\ A {\bf 48} 2574.
\bibitem{textbk} Cohen--Tannoudji Diu C B and Laloe F 1977
 {\it  M\'ecanique Quantique} (Paris: Herman) p.570.
\bibitem{arvroz} Arvieu R and Rozmej P 1994 {\it Phys.~Rev.} A {\bf 50}
4376.
\bibitem{arvroz1} Arvieu R and Rozmej P 1995 {\it Phys.~Rev.} A {\bf 51}
104.
\bibitem{rozarv} Rozmej P and Arvieu R 1996 {\it J.~Phys.} B {\bf
29}1339.
\bibitem{rozarv1} Rozmej P and Arvieu R 1996 {\it Acta Phys. Polon.}
 B {\bf 27} 581.
\bibitem{rozbarv} Rozmej P Berej W and Arvieu R 
 {em invited talk at XXXI Zakopane School of Physics, September 3--11, 
  1996}, $[$to appear in {\it Acta Phys. Polon. A} (1997)$]$.
\bibitem{mikhajlov} Mikhajlov V V 1973 {\it Izv.\ Akad. Nauk SSSR, Ser.
Fiz.}
 {\bf 37} 2230 (1974 {\it Bull. Acad. Sci. USSR, Phys. Ser.} {\bf 37}
 187).
\bibitem{shore} Shore B W and Knight P L 1993 {\it J.\ of Modern Optics}
 {\bf 40} 1195.
\bibitem{phoen} Phoenix S J D and Knight P L 1990 {\it J.\ Opt.\ Soc.\
 Am.} B {\bf 7} 116. 
   
\end{thebibliography}
\end{document}